\documentclass[a4paper,12pt]{article}
\usepackage{amsmath}
\usepackage{amssymb}
\usepackage{amsthm}

\theoremstyle{plain}

\theoremstyle{definition}

\theoremstyle{remark}

\numberwithin{equation}{section}

\title{Lax forms of the $q$-Painlev\'e equations}
\author{Mikio Murata\\
Department of Physics and Mathematics, \\
Aoyama Gakuin University, \\
5-10-1 Fuchinobe Sagamihara-shi, \\
Kanagawa 229-8558, Japan
}
\date{}

\begin{document}
\maketitle
\begin{abstract}
All $q$-Painlev\'e equations which are obtained from the $q$-analog of the sixth 
Painlev\'e equation are expressed  
in a Lax formalism. 
They are characterized by the data of the associated linear 
$q$-difference equations. 
The degeneration pattern from the $q$-Painlev\'e equation of type $A_2$ 
is also presented.
\end{abstract}



\section{Introduction}

Discrete Painlev\'e equations are studied from various points of view as integrable systems (\cite{RGH}). 
They are discrete equations which are reduced to the Painlev\'e equations in suitable limiting process, and moreover, which pass the singularity confinement test. 
Passing this test can be thought of as a difference analogue of the Painlev\'e property. 
The singularity confinement test has been proposed by Grammaticos et al. as a criterion for the integrability of discrete dynamical systems (\cite{GRP}). 

Discrete Painlev\'e equations were classified on the basis of the types of rational surfaces connected to extended affine Weyl groups (\cite{S2,S3}). 
There are three types of discrete Painlev\'e equations: elliptic-difference, $q$-difference and difference. 
We especially give the list of $q$-Painlev\'e equations in the discrete Painlev\'e equations. See Table \ref{table:list}.
\begin{table}[htbp]
\begin{tabular}{|c||c|c|c|c|c|}
\hline
Abbrev.& $q$-$P(A_0^*)$ & $q$-$P(A_1)$& $q$-$P(A_2)$& $q$-$P(A_3)$& $q$-$P(A_4)$\\
\hline
Surface& $A_0^{(1)*}$ & $A_1^{(1)}$ & $A_2^{(1)}$ & $A_3^{(1)}$ & $A_4^{(1)}$ \\
\hline
Symmetry& $E_8^{(1)}$ & $E_7^{(1)}$ & $E_6^{(1)}$ & $D_5^{(1)}$ & $A_4^{(1)}$\\
\hline
\end{tabular}

\medskip

\begin{tabular}{|c|c|c|c|c|c|c}
\hline
$q$-$P(A_5)$& $q$-$P(A_5)^{\sharp}$& $q$-$P(A_6)$& $q$-$P(A_6)^{\sharp}$& $q$-$P(A_7)$&$q$-$P(A_7')$ \\
\hline
$A_5^{(1)}$ & $A_5^{(1)}$ & $A_6^{(1)}$ & $A_6^{(1)}$ & $A_7^{(1)}$ & $A_7^{(1)\prime}$ \\
\hline
{\tiny $(A_2+A_1)^{(1)}$ }&{\tiny  $(A_2+A_1)^{(1)}$} & {\tiny $(A_1+A_1)^{(1)}$ }&{\tiny  $(A_1+A_1)^{(1)}$ }& $A_1^{(1)}$ & $A_1^{(1)}$ \\
\hline
\end{tabular}
\caption{The $q$-Painlev\'e equations}
\label{table:list}
\end{table}
As is well-known, the sixth Painlev\'e equation yields the other five Painlev\'e equations by a process of coalescence. 
Among the $q$-Painlev\'e equations, the $q$-Painlev\'e equation of type $A_0^*$ ($q$-$P(A_0^*)$) is the most generic one because the other $q$-Painlev\'e equations can be obtained from this equation by limiting procedure. 
These equations are organized in a degeneration pattern obtained through coalescence. See Table \ref{table:deg}.
\begin{table}[htbp]
\begin{tabular}{ccccccccc}
$q$-$P(A_0^*)$ &$\to$ & $q$-$P(A_1)$ &$\to$ & $q$-$P(A_2)$ &$\to$ 
& $q$-$P(A_3)$ &$\to$ &
\end{tabular}

\medskip

\begin{tabular}{ccccccccc}
&$\to$&$q$-$P(A_4)$&$\to$ & $q$-$P(A_5)$ &$\to$ & $q$-$P(A_6)$ &$\to$ 
& $q$-$P(A_7)$ \\
&& &$ \searrow $ &  &$ \nearrow $ &   &$ \nearrow $&   \\
&& & & $q$-$P(A_5)^{\sharp}$ &$\to$ & $q$-$P(A_6)^{\sharp}$ &$\to$ 
& $q$-$P(A_7')$ 
\end{tabular}
\caption{Degeneration pattern for the $q$-Painlev\'e equations}
\label{table:deg}
\end{table}

Another important aspect of the Painlev\'e equations is their connection to the monodromy-preserving deformation of linear differential equations. 
The generalized Riemann problem was already studied for linear differential, difference and $q$-difference equations in the Birkhoff's paper, \cite{B}. 
Jimbo and Sakai studied the deformation of a $2 \times 2$ matrix system of $q$-difference equations and found the $q$-Painlev\'e equation of type $A_3$ ($q$-$P(A_3)$), which is commonly known as $q$-$\mathrm{P_{VI}}$ (\cite{JS}). 
Sakai also found a Lax form of the $q$-Painlev\'e equation of type $A_2$ ($q$-$P(A_2)$), a particular case of a $q$-Garnier system (\cite{S,S1}). 
Hay et al.\ found Lax forms of $q$-Painlev\'e equations, 
reductions from a Lax pair for a lattice modified KdV equation (\cite{HHJN}). 
However, Lax forms of a lot of $q$-Painlev\'e equations have not been 
obtained yet. 

In this paper, we present Lax pairs of all $q$-Painlev\'e 
equations which are obtained from $q$-$P(A_3)$.
In Section~\ref{sec:lax}, we illustrate the connection preserving deformation 
and derive $q$-$P(A_3)$. 
We also propose Lax pairs of $q$-$P(A_4)$, $q$-$P(A_5)$, $q$-$P(A_5)^{\sharp}$, 
$q$-$P(A_6)$, $q$-$P(A_6)^{\sharp}$, $q$-$P(A_7)$ and $q$-$P(A_7')$ .
In Section~\ref{sec:deg}, we give replacements of the parameters for the degeneration. 
The Lax form of $q$-$P(A_3)$ can be obtained from the Lax form of $q$-$P(A_2)$. 
In Section~\ref{sec:a2a3}, we give the Lax form of $q$-$P(A_2)$ and 
replacements of the parameters for the degeneration.

\section{Lax forms of $q$-Painlev\'e equations}\label{sec:lax}

\subsection{Derivation of $q$-$P(A_3)$}
In this section, we illustrate the connection preserving deformation 
and derive the $q$-Painlev\'e equation of type $A_3$.

Consider a $2 \times 2$ matrix system with polynomial coefficients
\begin{equation}
Y(qx,t)=A(x,t)Y(x,t).\label{eq:a3x}
\end{equation}
The connection preserving deformation of the linear $q$-difference equation, 
which is a discrete counterpart of monodromy preserving deformation, is equivalent to existence of a linear deformation equation whose coefficients are rational in $x$. 
We express the deformation equation in the form 
\begin{equation}
Y(x,qt)=B(x,t)Y(x,t).\label{eq:a3t}
\end{equation}
The compatibility condition for the systems (\ref{eq:a3x}) and 
(\ref{eq:a3t}) reads 
\begin{equation}\label{eq:a3cc}
A(x,qt)B(x,t)=B(qx,t)A(x,t).
\end{equation}
$q$-$P(A_3)$ can be obtained from the condition (\ref{eq:a3cc}).

We take $A(x,t)$ to be of the form
\begin{gather}
A(x,t)=A_0(t)+xA_1(t)+x^2A_2,\\
A_2=
\begin{pmatrix}
\kappa_1&0\\
0&\kappa_2
\end{pmatrix},\quad
\text{$A_0(t)$ has eigenvalues $\theta_1 t$, $\theta_2 t$,}\\ 
 \det A(x,t)=\kappa_1\kappa_2(x-a_1t)(x-a_2t)(x-a_3)(x-a_4).
\end{gather}
Here the parameters $\kappa_j$, $\theta_j$, $a_j$ are independent of $t$. 
We have
\begin{equation}
\kappa_1\kappa_2a_1a_2a_3a_4=\theta_1\theta_2.
\end{equation}
Define $y=y(t)$, $z_i=z_i(t)\ (i=1,2)$ by
\begin{equation}
A_{12}(y,t)=0, \quad A_{11}(y,t)=\kappa_1 z_1,\quad 
A_{22}(y,t)=\kappa_2 z_2,
\end{equation}
so that 
\begin{equation}
z_1z_2=\kappa_1\kappa_2(y-a_1t)(y-a_2t)(y-a_3)(y-a_4).
\end{equation}
The matrix $A(x,t)$ can be parametrized as 
\begin{equation}
A(x,t)=
\begin{pmatrix}
\kappa_1((x-y)(x-\alpha)+z_1)&
\kappa_2w(x-y)\\
\kappa_1w^{-1}(\gamma x+\delta)&
\kappa_2((x-y)(x-\beta)+z_2)
\end{pmatrix}.
\end{equation}
Here
\begin{align}
\begin{split}
\alpha&=\frac{1}{\kappa_1-\kappa_2}
[y^{-1}((\theta_1+\theta_2)t-\kappa_1z_1-\kappa_2z_2)\\
&\quad{}-\kappa_2((a_1+a_2)t+a_3+a_4-2y)],
\end{split}\\
\begin{split}
\beta&=\frac{1}{\kappa_1-\kappa_2}
[-y^{-1}((\theta_1+\theta_2)t-\kappa_1z_1-\kappa_2z_2)\\
&\quad{}+\kappa_1((a_1+a_2)t+a_3+a_4-2y)],
\end{split}\\
\begin{split}
\gamma&=z_1+z_2+(y+\alpha)(y+\beta)+(\alpha+\beta)y-a_1a_2t^2\\
&\quad{}-(a_1+a_2)(a_3+a_4)t-a_3a_4,
\end{split}\\
\delta&=y^{-1}(a_1a_2a_3a_4t^2-(\alpha y+z_1)(\beta y+z_2)).
\end{align}
The quantity $w=w(t)$ is related to the \lq gauge' freedom, 
and does not enter the final result for the $q$-$P(A_3)$. 
The matrix $B(x,t)$ is a rational function of the form
\begin{equation}
B(x,t)=\frac{x}{(x-a_1 qt)(x-a_2qt)}(xI+B_0(t)).
\end{equation}
The compatibility (\ref{eq:a3cc}) is equivalent to 
\begin{gather} 
A(a_iqt,qt)(a_iqtI+B_0(t))=0\quad (i=1,2),\label{eq:a3cc1}\\
(a_iqtI+B_0(t))A(a_it,t)=0\quad (i=1,2),\label{eq:a3cc2}\\
A_0(qt)B_0(t)=qB_0(t)A_0(t).\label{eq:a3cc3}
\end{gather}
Substituting the parametrization above, one obtains a set of 
$q$-difference equations. 
Let us use the notations $\bar{y}=y(qt)$ and so forth. 
Introduce $z$ by 
\begin{equation}
z_1=\frac{(y-a_1 t)(y-a_2 t)}{\kappa_1qz},\quad
z_2=\kappa_1qz(y-a_3)(y-a_4).
\end{equation}
Then the matrix $B_0(t)=(B_{ij})$ is parametrized as follows: 
\begin{align*}
B_{11}&=\frac{-\kappa_2q\bar{z}}{1-\kappa_2\bar{z}}
\left(-\beta+\frac{t(a_1+a_2)-y}{\kappa_2\bar{z}}\right),\\
B_{12}&=\frac{\kappa_2qw\bar{z}}{1-\kappa_2\bar{z}},\\
\begin{split}
B_{21}&=\frac{\kappa_1q\bar{z}}{w(1-\kappa_1q\bar{z})}
\left(a_1qt-\bar{\alpha}+\frac{a_2qt-\bar{y}}{\kappa_1q\bar{z}}\right)
\left(a_1t-\beta+\frac{a_2t-y}{\kappa_2\bar{z}}\right)\\
&=\frac{\kappa_1q\bar{z}}{w(1-\kappa_1q\bar{z})}
\left(a_2qt-\bar{\alpha}+\frac{a_1qt-\bar{y}}{\kappa_1q\bar{z}}\right)
\left(a_2t-\beta+\frac{a_1t-y}{\kappa_2\bar{z}}\right),
\end{split}\\
B_{22}&=\frac{-\kappa_1q\bar{z}}{1-\kappa_1q\bar{z}}
\left(-\bar{\alpha}+\frac{qt(a_1+a_2)-\bar{y}}{\kappa_1q\bar{z}}\right).
\end{align*}
Set further 
\begin{equation}
b_1=\frac{a_1a_2}{\theta_1},\quad
b_2=\frac{a_1a_2}{\theta_2},\quad
b_3=\frac{1}{\kappa_1q},\quad
b_4=\frac{1}{\kappa_2}.
\end{equation}
Equations (\ref{eq:a3cc1})--(\ref{eq:a3cc3}) are equivalent to
\begin{align}
\frac{y\bar{y}}{a_3a_4}
&=\frac{(\bar{z}-b_1t)(\bar{z}-b_2t)}{(\bar{z}-b_3)(\bar{z}-b_4)},\label{eq:a3y}\\
\frac{z\bar{z}}{b_3b_4}
&=\frac{(y-a_1t)(y-a_2t)}{(y-a_3)(y-a_4)},\label{eq:a3z}\\
\frac{\bar{w}}{w}&=\frac{b_4}{b_3}\frac{\bar{z}-b_3}{\bar{z}-b_4}.
\end{align}
We have a single constraint 
\begin{equation}
\frac{b_1b_2}{b_3b_4}=q\frac{a_1a_2}{a_3a_4}.
\end{equation}
$q$-$P(A_3)$ is (\ref{eq:a3y}) and (\ref{eq:a3z}).

\subsection{Lax form of $q$-$P(A_4)$}
Consider a $2 \times 2$ matrix system with polynomial coefficients
\begin{equation}
Y(qx,t)=A(x,t)Y(x,t).
\end{equation}
We express the deformation equation in the form 
\begin{equation}
Y(x,qt)=B(x,t)Y(x,t)
\end{equation}
and can express the $q$-Painlev\'e equation of type $A_4$ in the form 
\begin{equation}
A(x,qt)B(x,t)=B(qx,t)A(x,t)\label{eq:a4cc}
\end{equation}
by the compatibility of the deformation equation and the original linear 
$q$-differ\-ence equation. 

We take $A(x,t)$ to be of the form
\begin{gather}
A(x,t)=A_0(t)+xA_1(t)+x^2A_2,\\
A_2=
\begin{pmatrix}
\kappa_1&0\\
0&0
\end{pmatrix},\quad
\text{$A_0(t)$ has eigenvalues $\theta_1 t$, $\theta_2 t$,}\\
\det A(x,t)=\kappa_1\kappa_2(x-a_1t)(x-a_2t)(x-a_3).
\end{gather}
We have
\begin{equation}
-\kappa_1\kappa_2a_1a_2a_3=\theta_1\theta_2.
\end{equation}
The matrix $B(x,t)$ is a rational function of the form
\begin{equation}
B(x,t)=\frac{x}{(x-a_1 qt)(x-a_2qt)}(xI+B_0(t)).
\end{equation}
Define $y=y(t)$, $z_i=z_i(t)\ (i=1,2)$ by
\begin{equation}
A_{12}(y,t)=0, \quad A_{11}(y,t)=\kappa_1 z_1,\quad 
A_{22}(y,t)= z_2,
\end{equation}
so that 
\begin{equation}
z_1z_2=\kappa_2(y-a_1t)(y-a_2t)(y-a_3).
\end{equation}
The matrix $A(x,t)$ can be parametrized as 
\begin{equation}
A(x,t)=
\begin{pmatrix}
\kappa_1((x-y)(x-\alpha)+z_1)&
w(x-y)\\
\kappa_1w^{-1}(\gamma x+\delta)&
\kappa_2(x-y)+z_2
\end{pmatrix}.
\end{equation}
Here
\begin{align}
\alpha&=\frac{1}{\kappa_1}
[y^{-1}((\theta_1+\theta_2)t-\kappa_1z_1-z_2)+\kappa_2],\\
\gamma&=z_2-\kappa_2((2y+\alpha)-(a_1+a_2)t-a_3),\\
\delta&=y^{-1}(-\kappa_2a_1a_2a_3t^2-(\alpha y+z_1)(-\kappa_2y+z_2)).
\end{align}
Introduce $z$ by 
\begin{equation}
z_1=\frac{(y-a_1 t)(y-a_2 t)}{\kappa_1qz},\quad
z_2=\kappa_1\kappa_2qz(y-a_3).
\end{equation}
Then the matrix $B_0(t)=(B_{ij})$ is parametrized as follows: 
\begin{align}
B_{11}&=-q\bar{z}
\left(\kappa_2+\frac{t(a_1+a_2)-y}{\bar{z}}\right),\\
B_{12}&=qw\bar{z},\\
\begin{split}
B_{21}&=\frac{\kappa_1q\bar{z}}{w(1-\kappa_1q\bar{z})}
\left(a_1qt-\bar{\alpha}+\frac{a_2qt-\bar{y}}{\kappa_1q\bar{z}}\right)
\left(\kappa_2+\frac{a_2t-y}{\bar{z}}\right)\\
&=\frac{\kappa_1q\bar{z}}{w(1-\kappa_1q\bar{z})}
\left(a_2qt-\bar{\alpha}+\frac{a_1qt-\bar{y}}{\kappa_1q\bar{z}}\right)
\left(\kappa_2+\frac{a_1t-y}{\bar{z}}\right),
\end{split}\\
B_{22}&=\frac{-\kappa_1q\bar{z}}{1-\kappa_1q\bar{z}}
\left(-\bar{\alpha}+\frac{qt(a_1+a_2)-\bar{y}}{\kappa_1q\bar{z}}\right).
\end{align}
Set further 
\begin{equation}
b_1=\frac{a_1a_2}{\theta_1},\quad
b_2=\frac{a_1a_2}{\theta_2},\quad
b_3=\frac{1}{\kappa_1q},\quad
a_4=-\kappa_2.
\end{equation}
Equation (\ref{eq:a4cc})  are equivalent to
\begin{align}
\frac{y\bar{y}}{a_3a_4}
&=-\frac{(\bar{z}-b_1t)(\bar{z}-b_2t)}
{\bar{z}-b_3},\label{eq:a4y}\\
\frac{z\bar{z}}{b_3}
&=-\frac{(y-a_1t)(y-a_2t)}{a_4(y-a_3)},\label{eq:a4z}\\
\frac{\bar{w}}{w}
&=-\frac{\bar{z}}{b_3}+1.
\end{align}
We have a constraint 
\begin{equation}
\frac{b_1b_2}{b_3}=q\frac{a_1a_2}{a_3a_4}.
\end{equation}
$q$-$P(A_4)$ is (\ref{eq:a4y}) and (\ref{eq:a4z}).

\subsection{Lax form of $q$-$P(A_5)$}
Consider a $2 \times 2$ matrix system with polynomial coefficients
\begin{equation}
Y(qx,t)=A(x,t)Y(x,t).
\end{equation}
We express the deformation equation
\begin{equation}
Y(x,qt)=B(x,t)Y(x,t)
\end{equation}
and can express the $q$-Painlev\'e equation of type $A_5$ in the form
\begin{equation}
A(x,qt)B(x,t)=B(qx,t)A(x,t)\label{eq:a5cc}
\end{equation}
by the compatibility of the deformation equation and the original linear 
$q$-differ\-ence equation. 

We take $A(x,t)$ to be of the form
\begin{gather}
A(x,t)=A_0(t)+xA_1(t)+x^2A_2,\\
A_2=
\begin{pmatrix}
\kappa_1&0\\
0&0
\end{pmatrix},\quad
\text{$A_0(t)$ has eigenvalues $\theta_1 t$, $0$,}\\
\det A(x,t)=\kappa_1\kappa_2x(x-a_1t)(x-a_2t).
\end{gather}
The matrix $B(x,t)$ is a rational function of the form
\begin{equation}
B(x,t)=\frac{x}{(x-a_1 qt)(x-a_2qt)}(xI+B_0(t)).
\end{equation}
Define $y=y(t)$, $z_i=z_i(t)\ (i=1,2)$ by
\begin{equation}
A_{12}(y,t)=0, \quad A_{11}(y,t)=\kappa_1 z_1,\quad 
A_{22}(y,t)= z_2,
\end{equation}
so that 
\begin{equation}
z_1z_2=\kappa_2y(y-a_1t)(y-a_2t).
\end{equation}
The matrix $A(x,t)$ can be parametrized as 
\begin{equation}
A(x,t)=
\begin{pmatrix}
\kappa_1((x-y)(x-\alpha)+z_1)&
w(x-y)\\
\kappa_1w^{-1}(\gamma x+\delta)&
\kappa_2(x-y)+z_2
\end{pmatrix}.
\end{equation}
Here
\begin{align}
\alpha&=\frac{1}{\kappa_1}
[y^{-1}(\theta_1t-\kappa_1z_1-z_2)+\kappa_2],\\
\gamma&=z_2-\kappa_2(2y+\alpha- t(a_1+a_2)),\\
\delta&=-y^{-1}(\alpha y+z_1)(-\kappa_2 y+z_2).
\end{align}
Introduce $z$ by 
\begin{equation}
z_1=\frac{(y-a_1 t)(y-a_2 t)}{\kappa_1qz},\quad
z_2=\kappa_1\kappa_2qzy.
\end{equation}
Then the matrix $B_0(t)=(B_{ij})$ is parametrized as follows: 
\begin{align}
B_{11}&=-q\bar{z}
\left(\kappa_2+\frac{t(a_1+a_2)-y}{\bar{z}}\right),\\
B_{12}&=qw\bar{z},\\
\begin{split}
B_{21}&=\frac{\kappa_1q\bar{z}}{w(1-\kappa_1q\bar{z})}
\left(a_1qt-\bar{\alpha}+\frac{a_2qt-\bar{y}}{\kappa_1q\bar{z}}\right)
\left(\kappa_2+\frac{a_2t-y}{\bar{z}}\right)\\
&=\frac{\kappa_1q\bar{z}}{w(1-\kappa_1q\bar{z})}
\left(a_2qt-\bar{\alpha}+\frac{a_1qt-\bar{y}}{\kappa_1q\bar{z}}\right)
\left(\kappa_2+\frac{a_1t-y}{\bar{z}}\right),
\end{split}\\
B_{22}&=\frac{-\kappa_1q\bar{z}}{1-\kappa_1q\bar{z}}
\left(-\bar{\alpha}+\frac{qt(a_1+a_2)-\bar{y}}{\kappa_1q\bar{z}}\right).
\end{align}
Set further 
\begin{equation}
b_1=\frac{a_1a_2}{\theta_1},\quad
b_2=-\frac{\theta_1}{\kappa_1\kappa_2},\quad
b_3=\frac{1}{\kappa_1q},\quad
a_4=-\kappa_2.
\end{equation}
Equation (\ref{eq:a5cc}) are equivalent to
\begin{align}
\frac{y\bar{y}}{a_3a_4}
&=\frac{b_2t( \bar{z}-b_1t)}
{\bar{z}-b_3},\label{eq:a5y}\\
\frac{z\bar{z}}{b_3}
&=-\frac{(y-a_1t)(y-a_2t)}{a_4y},\label{eq:a5z}\\
\frac{\bar{w}}{w}
&=-\frac{\bar{z}}{b_3}+ 1.
\end{align}
We have a constraint 
\begin{equation}
\frac{b_1b_2}{b_3}=q\frac{a_1a_2}{a_4}.
\end{equation}
$q$-$P(A_5)$ is (\ref{eq:a5y}) and (\ref{eq:a5z}).

\subsection{Lax form of $q$-$P(A_5)^{\sharp}$}
Consider a $2 \times 2$ matrix system with polynomial coefficients
\begin{equation}
Y(qx,t)=A(x,t)Y(x,t).
\end{equation}
We express the deformation equation 
\begin{equation}
Y(x,qt)=B(x,t)Y(x,t)
\end{equation}
and can express the $q$-Painlev\'e equation of type ${A_5}^{\sharp}$ in the form 
\begin{equation}
A(x,qt)B(x,t)=B(qx,t)A(x,t)\label{eq:a5scc}
\end{equation}
by the compatibility of the deformation equation and the original linear 
$q$-differ\-ence equation. 

We take $A(x,t)$ to be of the form
\begin{gather}
A(x,t)=A_0(t)+xA_1(t)+x^2A_2,\\
A_2=
\begin{pmatrix}
\kappa_1&0\\
0&0
\end{pmatrix},\quad
\text{$A_0(t)$ has eigenvalues $\theta_1 t$, $0$,}\\
\det A(x,t)=\kappa_1\kappa_2x(x-a_1t)(x-a_3).
\end{gather}
The matrix $B(x,t)$ is a rational function of the form
\begin{equation}
B(x,t)=\frac{1}{x-a_1 qt}(xI+B_0(t)).
\end{equation}
Define $y=y(t)$, $z_i=z_i(t)\ (i=1,2)$ by
\begin{equation}
A_{12}(y,t)=0, \quad A_{11}(y,t)=\kappa_1 z_1,\quad 
A_{22}(y,t)= z_2,
\end{equation}
so that 
\begin{equation}
z_1z_2=\kappa_2y(y-a_1t)(y-a_3).
\end{equation}
The matrix $A(x,t)$ can be parametrized as 
\begin{equation}
A(x,t)=
\begin{pmatrix}
\kappa_1((x-y)(x-\alpha)+z_1)&
w(x-y)\\
\kappa_1w^{-1}(\gamma x+\delta)&
\kappa_2(x-y)+z_2
\end{pmatrix}.
\end{equation}
Here
\begin{align}
\alpha&=\frac{1}{\kappa_1}
[y^{-1}(\theta_1t-\kappa_1z_1-z_2)+\kappa_2],\\
\gamma&=z_2-\kappa_2(2y+\alpha- a_1t-a_3),\\
\delta&=-y^{-1}(\alpha y+z_1)(-\kappa_2y+z_2).
\end{align}
Introduce $z$ by 
\begin{equation}
z_1=\frac{y(y-a_1 t)}{\kappa_1qz},\quad
z_2=\kappa_1\kappa_2qz(y-a_3).
\end{equation}
Then the matrix $B_0(t)=(B_{ij})$ is parametrized as follows: 
\begin{align}
B_{11}&=-q\bar{z}
\left(\kappa_2+\frac{a_1t-y}{\bar{z}}\right),\\
B_{12}&=qw\bar{z},\\
\begin{split}
B_{21}&=\frac{\kappa_1q\bar{z}}{w(1-\kappa_1q\bar{z})}
\left(a_1qt-\bar{\alpha}-\frac{\bar{y}}{\kappa_1q\bar{z}}\right)
\left(\kappa_2-\frac{y}{\bar{z}}\right)\\
&=\frac{\kappa_1q\bar{z}}{w(1-\kappa_1q\bar{z})}
\left(-\bar{\alpha}+\frac{a_1qt-\bar{y}}{\kappa_1q\bar{z}}\right)
\left(\kappa_2+\frac{a_1t-y}{\bar{z}}\right),
\end{split}\\
B_{22}&=\frac{-\kappa_1q\bar{z}}{1-\kappa_1q\bar{z}}
\left(-\bar{\alpha}+\frac{a_1qt-\bar{y}}{\kappa_1q\bar{z}}\right).
\end{align}
Set further 
\begin{equation}
b_1=\frac{a_1}{\theta_1},\quad
b_2=-\frac{\theta_1}{\kappa_1\kappa_2a_3},\quad
b_3=\frac{1}{\kappa_1q},\quad
a_4=-\kappa_2.
\end{equation}
Equation (\ref{eq:a5scc}) are equivalent to
\begin{align}
\frac{y\bar{y}}{a_3a_4}
&=-\frac{\bar{z}(\bar{z}-b_2t)}
{\bar{z}-b_3},\label{eq:a5sy}\\
\frac{z\bar{z}}{b_3}
&=-\frac{y(y-a_1t)}{a_4(y-a_3)},\label{eq:a5sz}\\
\frac{\bar{w}}{w}
&=-\frac{\bar{z}}{b_3}+ 1.
\end{align}
We have a constraint 
\begin{equation}
\frac{b_1b_2}{b_3}=q\frac{a_1}{a_3a_4}.
\end{equation}
$q$-$P(A_5)^{\sharp}$ is (\ref{eq:a5sy}) and (\ref{eq:a5sz}).

\subsection{Lax form of $q$-$P(A_6)$}
Consider a $2 \times 2$ matrix system with polynomial coefficients
\begin{equation}
Y(qx,t)=A(x,t)Y(x,t).
\end{equation}
We express the deformation equation 
\begin{equation}
Y(x,qt)=B(x,t)Y(x,t)
\end{equation}
and can express the $q$-Painlev\'e equation type $A_6$ in the form 
\begin{equation}
A(x,qt)B(x,t)=B(qx,t)A(x,t)\label{eq:a6cc}
\end{equation}
by the compatibility of the deformation equation and the original linear 
$q$-differ\-ence equation. 

We take $A(x,t)$ to be of the form
\begin{gather}
A(x,t)=A_0(t)+xA_1(t)+x^2A_2,\\
A_2=
\begin{pmatrix}
\kappa_1&0\\
0&0
\end{pmatrix},\quad
\text{$A_0(t)$ has eigenvalues $\theta_1 t$, $0$,}\\
\det A(x,t)=\kappa_1\kappa_2x^2(x-a_1t).
\end{gather}
The matrix $B(x,t)$ is a rational function of the form
\begin{equation}
B(x,t)=\frac{1}{x-a_1 qt}(xI+B_0(t)).
\end{equation}
Define $y=y(t)$, $z_i=z_i(t)\ (i=1,2)$ by
\begin{equation}
A_{12}(y,t)=0, \quad A_{11}(y,t)=\kappa_1 z_1,\quad 
A_{22}(y,t)= z_2,
\end{equation}
so that 
\begin{equation}
z_1z_2=\kappa_2y^2(y-a_1t).
\end{equation}
The matrix $A(x,t)$ can be parametrized as 
\begin{equation}
A(x,t)=
\begin{pmatrix}
\kappa_1((x-y)(x-\alpha)+z_1)&
w(x-y)\\
\kappa_1w^{-1}(\gamma x+\delta)&
\kappa_2(x-y)+z_2
\end{pmatrix}.
\end{equation}
Here
\begin{align}
\alpha&=\frac{1}{\kappa_1}
[y^{-1}(\theta_1 t-\kappa_1z_1-z_2)+\kappa_2],\\
\gamma&=z_2-\kappa_2(2y+\alpha-a_1t),\\
\delta&=-y^{-1}(\alpha y+z_1)(-\kappa_2 y+z_2).
\end{align}
Introduce $z$ by 
\begin{equation}
z_1=\frac{y(y-a_1 t)}{\kappa_1qz},\quad
z_2=\kappa_1\kappa_2qzy.
\end{equation}
Then the matrix $B_0(t)=(B_{ij})$ is parametrized as follows: 
\begin{align}
B_{11}&=-q\bar{z}
\left(\kappa_2+\frac{a_1t-y}{\bar{z}}\right),\\
B_{12}&=qw\bar{z},\\
\begin{split}
B_{21}&=\frac{\kappa_1q\bar{z}}{w(1-\kappa_1q\bar{z})}
\left(a_1qt-\bar{\alpha}-\frac{\bar{y}}{\kappa_1q\bar{z}}\right)
\left(\kappa_2-\frac{y}{\bar{z}}\right)\\
&=\frac{\kappa_1q\bar{z}}{w(1-\kappa_1q\bar{z})}
\left(-\bar{\alpha}+\frac{a_1qt-\bar{y}}{\kappa_1q\bar{z}}\right)
\left(\kappa_2+\frac{a_1t-y}{\bar{z}}\right).
\end{split}\\
B_{22}&=\frac{-\kappa_1q\bar{z}}{1-\kappa_1q\bar{z}}
\left(-\bar{\alpha}+\frac{a_1qt-\bar{y}}{\kappa_1q\bar{z}}\right).
\end{align}
Set further 
\begin{equation}
b_1=\frac{a_1}{\theta_1},\quad
b_2=-\frac{\theta_1}{\kappa_1\kappa_2},\quad
b_3=\frac{1}{\kappa_1q},\quad
b_4=-\kappa_2.
\end{equation}
Equation (\ref{eq:a6cc})  are equivalent to
\begin{align}
\frac{y\bar{y}}{a_4}
&=\frac{b_2t\bar{z}}
{\bar{z}-b_3},\label{eq:a6y}\\
\frac{z\bar{z}}{b_3}
&=-\frac{y(y-a_1t)}{a_4 y},\label{eq:a6z}\\
\frac{\bar{w}}{w}
&=-\frac{\bar{z}}{b_3}+ 1.
\end{align}
We have a constraint 
\begin{equation}
\frac{b_1b_2}{b_3}=q\frac{a_1}{a_4}.
\end{equation}
$q$-$P(A_6)$ is (\ref{eq:a6y}) and (\ref{eq:a6z}).

\subsection{Lax form of $q$-$P(A_6)^{\sharp}$}
Consider a $2 \times 2$ matrix system with polynomial coefficients
\begin{equation}
Y(qx,t)=A(x,t)Y(x,t).
\end{equation}
We express the deformation equation
\begin{equation}
Y(x,qt)=B(x,t)Y(x,t)
\end{equation}
and can express the $q$-Painlev\'e equation of type ${A_6}^{\sharp}$ in the form 
\begin{equation}
A(x,qt)B(x,t)=B(qx,t)A(x,t)\label{eq:a6scc}
\end{equation}
by the compatibility of the deformation equation and the original linear 
$q$-differ\-ence equation. 

We take $A(x,t)$ to be of the form
\begin{gather}
A(x,t)=A_0(t)+xA_1(t)+x^2A_2,\\
A_2=
\begin{pmatrix}
\kappa_1&0\\
0&0
\end{pmatrix},\quad
\text{$A_0(t)$ has eigenvalues $\theta_1 t$, $0$,}\\
\det A(x,t)=\kappa_1\kappa_2x^2(x-a_3),
\end{gather}
The matrix $B(x,t)$ is a rational function of the form
\begin{equation}
B(x,t)=\frac{1}{x}(xI+B_0(t)).
\end{equation}
Define $y=y(t)$, $z_i=z_i(t)\ (i=1,2)$ by
\begin{equation}
A_{12}(y,t)=0, \quad A_{11}(y,t)=\kappa_1 z_1,\quad 
A_{22}(y,t)= z_2,
\end{equation}
so that 
\begin{equation}
z_1z_2=\kappa_2y^2(y-a_3).
\end{equation}
The matrix $A(x,t)$ can be parametrized as 
\begin{equation}
A(x,t)=
\begin{pmatrix}
\kappa_1((x-y)(x-\alpha)+z_1)&
w(x-y)\\
\kappa_1w^{-1}(\gamma x+\delta)&
\kappa_2(x-y)+z_2
\end{pmatrix}.
\end{equation}
Here
\begin{align}
\alpha&=\frac{1}{\kappa_1}
[y^{-1}(\theta_1 t-\kappa_1z_1-z_2)+\kappa_2],\\
\gamma&=z_2-\kappa_2(2y+\alpha-a_3),\\
\delta&=-y^{-1}(\alpha y+z_1)(-\kappa_2 y+z_2).
\end{align}
Introduce $z$ by 
\begin{equation}
z_1=\frac{y^2}{\kappa_1qz},\quad
z_2=\kappa_1\kappa_2qz(y-a_3).
\end{equation}
Then the matrix $B_0(t)=(B_{ij})$ is parametrized as follows: 
\begin{align}
B_{11}&=-q\bar{z}
\left(\kappa_2-\frac{y}{\kappa_2\bar{z}}\right),\\
B_{12}&=qw\bar{z},\\
B_{21}&=\frac{\kappa_1q}{w(1-\kappa_1q\bar{z})}
\left(-\bar{\alpha}-\frac{\bar{y}}{\kappa_1q\bar{z}}\right)
\left(\kappa_2-\frac{y}{\bar{z}}\right),\\
B_{22}&=\frac{-\kappa_1q\bar{z}}{1-\kappa_1q\bar{z}}
\left(-\bar{\alpha}-\frac{\bar{y}}{\kappa_1q\bar{z}}\right).
\end{align}
Set further 
\begin{equation}
b_1=\frac{1}{\theta_1} ,\quad
b_2=-\frac{\theta_1}{\kappa_1\kappa_2a_3},\quad
b_3=\frac{1}{\kappa_1q},\quad
a_4=-\kappa_2.
\end{equation}
Equation (\ref{eq:a6scc}) are equivalent to
\begin{align}
\frac{y\bar{y}}{a_3a_4}
&=-\frac{\bar{z}(\bar{z}-b_2t)}
{\bar{z}-b_3},\label{eq:a6sy}\\
\frac{z\bar{z}}{b_3}
&=-\frac{y^2}{a_4(y-a_3)},\label{eq:a6sz}\\
\frac{\bar{w}}{w}
&=-\frac{\bar{z}}{b_3}+ 1.
\end{align}
We have a constraint 
\begin{equation}
\frac{b_1b_2}{b_3}=q\frac{1}{a_3a_4}.
\end{equation}
$q$-$P(A_6)^{\sharp}$ is (\ref{eq:a6sy}) and (\ref{eq:a6sz}).

\subsection{Lax form of $q$-$P(A_7)$}
Consider a $2 \times 2$ matrix system with polynomial coefficients
\begin{equation}
Y(qx,t)=A(x,t)Y(x,t).
\end{equation}
We express the deformation equation
\begin{equation}
Y(x,qt)=B(x,t)Y(x,t)
\end{equation}
and can express the $q$-Painlev\'e equation of type $A_7$ in the form 
\begin{equation}
A(x,qt)B(x,t)=B(qx,t)A(x,t)\label{eq:a7cc}
\end{equation}
by the compatibility of the deformation equation and the original linear 
$q$-differ\-ence equation. 

We take $A(x,t)$ to be of the form
\begin{gather}
A(x,t)=A_0(t)+xA_1(t)+x^2A_2,\\
A_2=
\begin{pmatrix}
\kappa_1&0\\
0&0
\end{pmatrix},\quad
\text{$A_0(t)$ has eigenvalues $\theta_1 t$, $0$,}\\
\det A(x,t)=\kappa_1\kappa_2x^3.
\end{gather}
The matrix $B(x,t)$ is a rational function of the form
\begin{equation}
B(x,t)=\frac{1}{x}(xI+B_0(t)).
\end{equation}
Define $y=y(t)$, $z_i=z_i(t)\ (i=1,2)$ by
\begin{equation}
A_{12}(y,t)=0, \quad A_{11}(y,t)=\kappa_1 z_1,\quad 
A_{22}(y,t)= z_2,
\end{equation}
so that 
\begin{equation}
z_1z_2=\kappa_2y^3.
\end{equation}
The matrix $A(x,t)$ can be parametrized as 
\begin{equation}
A(x,t)=
\begin{pmatrix}
\kappa_1((x-y)(x-\alpha)+z_1)&
w(x-y)\\
\kappa_1w^{-1}(\gamma x+\delta)&
\kappa_2(x-y)+z_2
\end{pmatrix}.
\end{equation}
Here
\begin{align}
\alpha&=\frac{1}{\kappa_1}
[y^{-1}(\theta_1 t-\kappa_1z_1-z_2)+\kappa_2],\\
\gamma&=z_2-\kappa_2(2y+\alpha),\\
\delta&=-y^{-1}(\alpha y+z_1)(\beta y+z_2).
\end{align}
Introduce $z$ by 
\begin{equation}
z_1=\frac{y^2}{\kappa_1qz},\quad
z_2=\kappa_1\kappa_2qyz.
\end{equation}
Then the matrix $B_0(t)=(B_{ij})$ is parametrized as follows: 
\begin{align}
B_{11}&=-q\bar{z}\left(\kappa_2-\frac{y}{\bar{z}}\right),\\
B_{12}&=qw\bar{z},\\
B_{21}&=\frac{\kappa_1q\bar{z}}{w(1-\kappa_1q\bar{z})}
\left(-\bar{\alpha}-\frac{\bar{y}}{\kappa_1q\bar{z}}\right)
\left(\kappa_2-\frac{y}{\bar{z}}\right),\\
B_{22}&=\frac{-\kappa_1q\bar{z}}{1-\kappa_1q\bar{z}}
\left(-\bar{\alpha}-\frac{\bar{y}}{\kappa_1q\bar{z}}\right).
\end{align}
Set further 
\begin{equation}
b_1=\frac{1}{\theta_1} ,\quad
b_2=-\frac{\theta_1}{\kappa_1\kappa_2},\quad
b_3=\frac{1}{\kappa_1q},\quad
a_4=-\kappa_2.
\end{equation}
Equation (\ref{eq:a7cc}) are equivalent to
\begin{align}
\frac{y\bar{y}}{a_4}
&=\frac{b_2t\bar{z}}
{\bar{z}-b_3},\label{eq:a7y}\\
\frac{z\bar{z}}{b_3}
&=-\frac{y}{a_4},\label{eq:a7z}\\
\frac{\bar{w}}{w}
&=-\frac{\bar{z}}{b_3}+ 1.
\end{align}
We have a constraint 
\begin{equation}
\frac{b_1b_2}{b_3}=q\frac{1}{a_4}.
\end{equation}
$q$-$P(A_7)$ is (\ref{eq:a7y}) and (\ref{eq:a7z}).

\subsection{Lax form of $q$-$P(A_7^{\prime})$}
Consider a $2 \times 2$ matrix system with polynomial coefficients
\begin{equation}
Y(qx,t)=A(x,t)Y(x,t).
\end{equation}
We express the deformation equation
\begin{equation}
Y(x,qt)=B(x,t)Y(x,t)
\end{equation}
and can express the $q$-Painlev\'e equation of type $A_7^{\prime}$ in the form 
\begin{equation}
A(x,qt)B(x,t)=B(qx,t)A(x,t)\label{eq:a7pcc}
\end{equation}
by the compatibility of the deformation equation and the original linear 
$q$-differ\-ence equation. 

We take $A(x,t)$ to be of the form
\begin{gather}
A(x,t)=A_0(t)+xA_1(t)+x^2A_2,\\
A_2=
\begin{pmatrix}
\kappa_1&0\\
0&0
\end{pmatrix},\quad
\text{$A_0(t)$ has eigenvalues $\theta_1 t$, $0$,}\\
\det A(x,t)=\kappa_1\kappa_2x^2.
\end{gather}
The matrix $B(x,t)$ is a rational function of the form
\begin{equation}
B(x,t)=\frac{1}{x}(xI+B_0(t)).
\end{equation}
Define $y=y(t)$, $z_i=z_i(t)\ (i=1,2)$ by
\begin{equation}
A_{12}(y,t)=0, \quad A_{11}(y,t)=\kappa_1 z_1,\quad 
A_{22}(y,t)= z_2,
\end{equation}
so that 
\begin{equation}
z_1z_2=\kappa_2y^2.
\end{equation}
The matrix $A(x,t)$ can be parametrized as 
\begin{equation}
A(x,t)=
\begin{pmatrix}
\kappa_1((x-y)(x-\alpha)+z_1)&
w(x-y)\\
\kappa_1w^{-1}(\gamma x+\delta)&
 z_2
\end{pmatrix}.
\end{equation}
Here
\begin{align}
\alpha&=\frac{1}{\kappa_1}
y^{-1}(\theta_1 t-\kappa_1z_1-z_2),\\
\gamma&=z_2+\kappa_2,\\
\delta&=-y^{-1}z_2(\alpha y+z_1).
\end{align}
Introduce $z$ by 
\begin{equation}
z_1=\frac{y^2}{\kappa_1qz},\quad
z_2=\kappa_1\kappa_2qz.
\end{equation}
Then the matrix $B_0(t)=(B_{ij})$ is parametrized as follows: 
\begin{align}
B_{11}&=qy,\\
B_{12}&=qw\bar{z},\\
B_{21}&=\frac{-\kappa_1qy}{w(1-\kappa_1q\bar{z})}
\left(-\bar{\alpha}-\frac{\bar{y}}{\kappa_1q\bar{z}}\right),\\
B_{22}&=\frac{-\kappa_1q\bar{z}}{1-\kappa_1q\bar{z}}
\left(-\bar{\alpha}-\frac{\bar{y}}{\kappa_1q\bar{z}}\right).
\end{align}
Set further 
\begin{equation}
b_1=\frac{1}{\theta_1},\quad
b_2=-\frac{\theta_1}{\kappa_1\kappa_2},\quad
b_3=\frac{1 }{\kappa_1q},\quad
b_4=-\kappa_2.
\end{equation}
Equation (\ref{eq:a7pcc}) are equivalent to
\begin{align}
\frac{y\bar{y}}{a_4}
&=-\frac{\bar{z}(\bar{z}-b_2t)}
{\bar{z}-b_3},\label{eq:a7py}\\
\frac{z\bar{z}}{b_3}
&=\frac{y^2}{a_4},\label{eq:a7pz}\\
\frac{\bar{w}}{w}
&=-\frac{\bar{z}}{b_3}+ 1.
\end{align}
We have a constraint 
\begin{equation}
\frac{b_1b_2}{b_3}=q\frac{1}{a_4}.
\end{equation}
$q$-$P(A_7^{\prime})$ is (\ref{eq:a7py}) and (\ref{eq:a7pz}).

\section{Degenerations}\label{sec:deg}
Some replacements of the parameters for the degenerations of $q$-Painlev\'e 
equations were given in the paper, \cite{RGTT}, for example.
In this section, we present the replacements of the parameters of the Lax formalisms.

Replace in $q$-$P(A_3)$, $t$ by $\varepsilon t$, 
$y$ by $\varepsilon y$, $z$ by $\varepsilon z$, 
$a_3$ by $\varepsilon a_3$, $a_4$ by $\varepsilon^{-1} a_4$
$b_3$ by $\varepsilon b_3$ and $b_4$ by $\varepsilon^{-1}$ 
and let $\varepsilon$ tend to zero. 
Then we obtain  $q$-$P(A_4)$: 
\begin{align}
\frac{y\bar{y}}{a_3a_4}
&=-\frac{(\bar{z}-b_1t)(\bar{z}-b_2t)}
{\bar{z}-b_3},\\
\frac{z\bar{z}}{b_3}
&=-\frac{(y-a_1t)(y-a_2t)}{a_4(y-a_3)}.
\end{align}
\begin{equation}
\frac{b_1b_2}{b_3}=q\frac{a_1a_2}{a_3a_4}
\end{equation}
For the sake of simplification of notation, 
the replacement and the succeeding limiting process 
will be written as follows: 
\begin{gather*}
t\to \varepsilon t\quad
y\to \varepsilon y,\quad
z\to \varepsilon z,\\
a_3\to \varepsilon a_3,\quad
a_4\to \varepsilon^{-1} a_4\quad
b_3\to \varepsilon b_3,\quad
b_4\to \varepsilon^{-1}.
\end{gather*}

By the use of notation as above, 
the degeneration from the Lax form of $q$-$P(A_3)$ to 
that of $q$-$P(A_4)$ is given by the following scheme: 

$q$-$P(A_3)$ to $q$-$P(A_4)$: 
\begin{gather*}
t\to \varepsilon t\quad
y\to \varepsilon y,\quad
z\to \varepsilon z,\\
a_3\to \varepsilon a_3,\quad
a_4\to \varepsilon^{-1} a_4\quad
b_3\to \varepsilon b_3,\quad
b_4\to \varepsilon^{-1},\\
x\to \varepsilon x,\quad
z_1\to \varepsilon^2 z_1,\quad
w\to \varepsilon^{-1} w,\quad
\kappa_1\to \varepsilon^{-1} \kappa_1,\quad
\kappa_2\to \varepsilon,\\
\alpha\to \varepsilon\alpha,\quad
\beta\to \varepsilon^{-1}\beta,\quad
\delta\to \varepsilon \delta,\\
Y(x,t)\to x^{\log_q \varepsilon}Y(x,t),\quad
A(x,t)\to \varepsilon A(x,t),\\
A_0(t)\to \varepsilon A_0(t),\quad
A_2\to \varepsilon^{-1}A_2,\quad
B_0(t)\to \varepsilon B_0(t),\\
B_{11}\to\varepsilon B_{11},\quad
B_{12}\to\varepsilon B_{12},\quad
B_{21}\to\varepsilon B_{21},\quad
B_{22}\to\varepsilon B_{22}.
\end{gather*}

$q$-$P(A_4)$ to $q$-$P(A_5)$: 
\begin{gather*}
a_3\to \varepsilon ,\quad
b_2\to \varepsilon^{-1} b_2,\quad
\theta_2\to \varepsilon .
\end{gather*}

$q$-$P(A_4)$ to $q$-$P(A_5)^{\sharp}$: 
\begin{gather*}
t\to\varepsilon t,\quad
a_1\to\varepsilon^{-1} a_1,\quad
a_2\to\varepsilon,\quad
b_1\to\varepsilon b_1,\quad
b_2\to\varepsilon^{-1} b_2,\\
\theta_1\to\varepsilon^{-1} \theta_1,\quad
\theta_2\to\varepsilon .
\end{gather*}

$q$-$P(A_5)$ to $q$-$P(A_6)$: 
\begin{gather*}
t\to \varepsilon t,\quad
a_1\to \varepsilon^{-1} a_1,\quad
a_2\to \varepsilon ,\quad 
b_1\to \varepsilon b_1,\quad 
b_2\to \varepsilon^{-1} b_2,\quad \theta_1\to \varepsilon^{-1} \theta_1.
\end{gather*}

$q$-$P(A_5)^{\sharp}$ to $q$-$P(A_6)$: 
\begin{gather*}
a_3\to \varepsilon ,\quad
b_2\to \varepsilon^{-1} b_2.
\end{gather*}

$q$-$P(A_5)^{\sharp}$ to $q$-$P(A_6)^{\sharp}$: 
\begin{gather*}
a_1\to \varepsilon ,\quad
b_1\to \varepsilon b_1.
\end{gather*}

$q$-$P(A_6)$ to $q$-$P(A_7)$: 
\begin{gather*}
a_1\to \varepsilon ,\quad
b_1\to \varepsilon b_1.
\end{gather*}

$q$-$P(A_6)^{\sharp}$ to $q$-$P(A_7)$: 
\begin{gather*}
a_3\to \varepsilon,\quad
b_2\to \varepsilon^{-1} b_2 .
\end{gather*}

$q$-$P(A_6)^{\sharp}$ to $q$-$P(A_7^{\prime})$: 
\begin{gather*}
a_3\to \varepsilon^{-1} ,\quad
a_4\to \varepsilon a_4,\quad
\kappa_2\to  \varepsilon \kappa_2.
\end{gather*}

\section{Degeneration from $q$-$P(A_2)$ to $q$-$P(A_3)$}\label{sec:a2a3}
The Lax form of $q$-$P(A_2)$ was given in the Sakai's paper, \cite{S1}.
This Lax form yields the Lax form of $q$-$P(A_3)$ by a process of coalescence.

\subsection{Lax form of $q$-$P(A_2)$}
In this subsection, we illustrate the Lax form of $q$-$P(A_2)$ in the paper, \cite{S1}.

Consider a $2 \times 2$ matrix system with polynomial coefficients
\begin{equation}
Y(qx,t)=A(x,t)Y(x,t).\label{eq:a2x}
\end{equation}
We express the deformation equation
\begin{equation}
Y(x,qt)=B(x,t)Y(x,t)\label{eq:a2t}
\end{equation}
and can express the $q$-Painlev\'e equation of type $A_2$ 
in the form 
\begin{equation}\label{eq:a2cc}
A(x,qt)B(x,t)=B(qx,t)A(x,t)
\end{equation}
by the compatibility of the deformation equation and the original linear 
$q$-differ\-ence equation. 

We take $A(x,t)$ to be of the form
\begin{gather}
A(x,t)=A_0(t)+xA_1(t)+x^2A_2(t)+x^3A_3,\\
A_3=
\begin{pmatrix}
\kappa_1&0\\
0&\kappa_2
\end{pmatrix},\quad
\text{$A_0(t)$ has eigenvalues $\theta_1t$, $\theta_2t$,}\\ 
 \det A(x,t)=\kappa_1\kappa_2(x-a_1)(x-a_2)(x-a_3)(x-a_4)(x-a_5t)(x-a_6t).
\end{gather}
We have
\begin{equation}
\kappa_1\kappa_2a_1a_2a_3a_4a_5a_6=\theta_1\theta_2.
\end{equation}
The matrix $B(x,t)$ is a rational function of the form
\begin{equation}
B(x,t)=\frac{x}{(x-a_5qt)(x-a_6qt)}(xI+B_0(t)).
\end{equation}
Define $\lambda=\lambda(t)$, $\mu=\mu(t)$ and $\tilde{\mu}=\tilde{\mu}(t)$  by
\begin{equation}
A_{12}(\lambda,t)=0, \quad A_{11}(\lambda,t)=\kappa_1 \tilde{\mu},\quad 
A_{22}(\lambda,t)=\kappa_2 \mu,
\end{equation}
so that 
\begin{equation}
\mu\tilde{\mu}
=\kappa_1\kappa_2(\lambda-a_1)(\lambda-a_2)(\lambda-a_3)(\lambda-a_4)(\lambda-a_5t)(\lambda-a_6t).
\end{equation}
The matrix $A(x,t)$ can be parametrized as 
\begin{equation}
A(x,t)=
\begin{pmatrix}
\kappa_1W(x,t)&\kappa_2 w L(x,t)\\
\kappa_1w^{-1}X(x,t)&\kappa_2Z(x,t)
\end{pmatrix},
\end{equation}
Here
\begin{align}
L(x,t)&=x-\lambda,\\
Z(x,t)&=(x-\lambda)(x^2+(\gamma+\lambda)x+\delta)+\mu,\\
W(x,t)&=(x-\lambda)(x^2+(-\gamma+\lambda-\sigma_1)x+\tilde{\delta})+\tilde{\mu},\\
X(x,t)&=\frac{W(x)Z(x)-\prod_{i=1}^6(x-a_i)}{L(x)},\\
\delta&=\frac{1}{\kappa_1-\kappa_2}\left[\kappa_1(2\lambda^2-\sigma_1\lambda+\sigma_2+\gamma(\gamma+\sigma_1))-\frac{1}{\lambda}(\kappa_1\tilde{\mu}+\kappa_2\mu-\theta_1-\theta_2)\right],\\
\tilde{\delta}&=\frac{1}{\kappa_1-\kappa_2}\left[-\kappa_2(2\lambda^2-\sigma_1\lambda+\sigma_2+\gamma(\gamma+\sigma_1))+\frac{1}{\lambda}(\kappa_1\tilde{\mu}+\kappa_2\mu-\theta_1-\theta_2)\right],\\
\tilde{\mu}&=\frac{1}{\mu}\prod_{i=1}^6(\lambda-a_i),\quad
\sigma_1=\sum_{i=1}^6a_i,\quad
\sigma_2=\sum_{i<j}a_ia_j.
\end{align}
If $q\kappa_1=\kappa_2$, then Equation (\ref{eq:a2cc}) are equivalent to
\begin{gather}
(\lambda-\underline{\nu})(\lambda-\nu)
=\frac{(\lambda-a_1)(\lambda-a_2)(\lambda-a_3)(\lambda-a_4)}
{(\lambda-a_5 t)(\lambda-a_6 t)},\label{eq:a2y}\\
\left(1-\frac{\nu}{\bar{\lambda}}\right)\left(1-\frac{\nu}{\lambda}\right)
=\frac{a_5a_6}{q}\frac{(\nu-a_1)(\nu-a_2)(\nu-a_3)(\nu-a_4)}
{(a_5a_6t+\theta_1/\kappa_2)(a_5a_6t+\theta_2/\kappa_2)},\label{eq:a2z}\\
\begin{split}
a_5a_6t\lambda\bar{\lambda}(a_1;a_2+a_3+a_4+\bar{\gamma}-\nu)
((a_5+a_6)t+\gamma+\nu)&\\
{}+
q(a_5a_6t\nu+\theta_1/\kappa_2)(a_5a_6t\nu+\theta_2/\kappa_2)&=0.
\end{split}
\end{gather}
$q$-$P(A_2)$ is (\ref{eq:a2y}) and (\ref{eq:a2z}).

\subsection{Degeneration}
In this subsection, we give replacements of the parameters for the degeneration. 
By the use of notation in the Section~\ref{sec:deg}, 
the degeneration from the Lax form of $q$-$P(A_2)$ to 
that of $q$-$P(A_3)$ is given by the following scheme: 
\begin{gather*}
\lambda\to\varepsilon y,\quad
\nu\to \varepsilon^{-1}z,\\
a_1\to \varepsilon a_3,\quad
a_2\to \varepsilon a_4,\quad
a_3\to -\varepsilon^{-1},\\
a_4\to -\varepsilon^{-1} q\kappa_2^{-1}\kappa_1,\quad
a_5\to \varepsilon a_1,\quad
a_6\to \varepsilon a_2,\\
x\to \varepsilon x,\quad
\mu\to\varepsilon z_2,\quad
\tilde{\mu}\to\varepsilon q {\kappa_2}^{-1}\kappa_1 z_1,\quad
\kappa_1\to \varepsilon^{-1} q^{-1}\kappa_2,\quad
\kappa_2\to \varepsilon^{-1} \kappa_2,\\
\gamma\to \varepsilon^{-1}+  \varepsilon\gamma_1+O(\varepsilon^2),\\
\begin{split}
\gamma_1&=\frac{1}{\kappa_1-\kappa_2}
[y^{-1}((\theta_1+\theta_2)t-\kappa_1z_1-\kappa_2 z_2)\\
&\quad{}-\kappa_2((a_1+a_2)t+a_3+a_4)+y(\kappa_1+\kappa_2)].
\end{split}
\end{gather*}

\section{Concluding remarks}
By the limiting procedure, we derived Lax forms of $q$-Painlev\'e 
equations which are obtained from $q$-$P(A_3)$. 
The degeneration scheme from the Lax forms of $q$-$P(A_2)$ is also given. 
However Lax forms of $q$-$P(A_0^*)$ and $q$-$P(A_1)$ do not appear today, 
the full degeneration pattern cannot be presented.  
An interesting future problem remains to find 
the relation between the Lax pairs in the paper, 
\cite{HHJN}, and our result.  

%$A_2$ has the eigenvalue $0$, and is not invertible, 

%The Painlev\'e equations except for the sixth Painlev\'e equation appear 
%as a deformation condition of the linear differential equations with irregular singularity.  

\subsection*{Acknowledgments}
The author expresses his sincere gratitude to Professor Hidetaka Sakai,  
who gave suggestions about ideas on this research.

\bibliographystyle{plain}

\begin{thebibliography}{10}

\bibitem{B}
G. D. Birkhoff, 
The generalized Riemann problem for linear differential equations and the allied problems for linear difference and $q$-difference equations, 
\textit{Proc.\ Amer.\ Acad.\ Arts Sci.}~\textbf{49} (1913), 
521--568.

\bibitem{GRP}
B. Grammticos, A. Ramani and V. G. Papageorgiou, 
Do integrable mappings have the Painlev\'e property?, 
\textit{Phys.\ Rev.\ Lett.}~\textbf{67} (1991), 1825--1828. 

\bibitem{HHJN}
M. Hay, J. Hietarinta, N. Joshi and F. Nijhoff, 
A Lax pair for a lattice modified KdV equation, reductions to $q$-Painlev\'e equations and associated Lax pairs, 
\textit{J.\ Phys.\ A: Math.\ Theor.}~\textbf{40} (2007), 
F61--F73.

\bibitem{JS}
M. Jimbo and H. Sakai, 
A $q$-analog of the sixth Painlev\'e equation, 
\textit{Lett. Math. Phys.}~\textbf{38} (1996), 145--154.

\bibitem{RGH}
A. Ramani, B. Grammaticos and J. Hietarinta, 
Discrete versions of the Painlev\'e equations, 
\textit{Phys.\ Rev.\ Lett.}~\textbf{67} (1991), 1829--1832.

\bibitem{RGTT}
A. Ramani, B. Grammaticos, T. Tamizhmani and K. M. Tamizhmani, 
Special function solutions of the discrete painlev\'e equations, 
\textit{Comput.\ Math.\ Appl.}~\textbf{42} (2001), 603--614.

\bibitem{S}
H. Sakai, 
A $q$-analog of the Garnier system, 
\textit{Funkcial.\ Ekvac.}~\textbf{48} (2005), 
273--297. 

\bibitem{S1}
H. Sakai, 
Lax form of the $q$-Painlev\'e equation associated with the $A_2^{(1)}$ 
surface, 
\textit{J.\ Phys.\ A: Math.\ Gen.}~\textbf{39} (2006), 
12203--12210. 

\bibitem{S2}
H. Sakai, 
Problem: discrete Painlev\'e equations and their Lax forms, 
\textit{RIMS K\^oky\^uroku Bessatsu}~\textbf{B2} (2007), 
195--208. 

\bibitem{S3}
H. Sakai, 
Rational surfaces associated with affine root systems and geometry of 
the Painlev\'e equations, 
\textit{Comm.\ Math.\ Phys.}~\textbf{220} (2001), 
165--229. 

\end{thebibliography}

\end{document}